\documentclass[pra,aps,twocolumn,superscriptaddress,showkeys,nofootinbib]{revtex4-1}
\usepackage[latin1]{inputenc}
\usepackage{amsmath,amsthm,amssymb,graphicx,subfigure,xcolor}
\usepackage{mathrsfs}
\usepackage{graphics,float}
\usepackage{epstopdf}
\usepackage{color}
\usepackage[unicode=true]{hyperref}
\usepackage{booktabs}
\usepackage{tabularx}
\usepackage{tabu}
\usepackage{multirow}
\allowdisplaybreaks[4]
\hypersetup{
     colorlinks=true,       		
     linkcolor=blue,          	
     citecolor=blue,            
     urlcolor=blue,           	
 }

\newtheorem*{theorem*}{Theorem}

\newtheorem*{corollary*}{Corollary}

\newtheorem*{lemma*}{Lemma}

\newtheorem*{proposition*}{Proposition}
\theoremstyle{definition}

\newtheorem*{definition*}{Definition}
\theoremstyle{remark}

\newtheorem*{remark*}{Remark}

\begin{document}
\title{Entanglement hierarchies in multipartite scenarios}
\author{Hui Li}
\author{Ting Gao}
\email{gaoting@hebtu.edu.cn}
\affiliation{School of Mathematical Sciences, Hebei Normal University, Shijiazhuang 050024, China}
\author{Fengli Yan}
\email{flyan@hebtu.edu.cn}
\affiliation{College of Physics, Hebei Key Laboratory of Photophysics Research and Application, Hebei Normal University, Shijiazhuang 050024, China}
\begin{abstract}
In this paper, we investigate the hierarchical structure of the $n$-partite quantum states. We present a whole set of hierarchical quantifications as a method of characterizing quantum states, which go beyond genuine multipartite entanglement measures and allow for fine identification among distinct entanglement contributions. This kind of quantifications, termed $k$-GM concurrence, can unambiguously classify entangled states into $(n-1)$ distinct classes from the perspective of $k$-nonseparability with $k$ running from $n$ down to 2, and comply with the axiomatic conditions of an entanglement measure. Compared to $k$-ME concurrence [\href{https://journals.aps.org/pra/abstract/10.1103/PhysRevA.86.062323} {Phys. Rev. A \textbf{86}, 062323 (2012)}], the hierarchical measures proposed by us embody advantages in distinguishing same class entangled state and measuring continuity. In addition, we establish the relation between $k$-ME concurrence and $k$-GM concurrence, and further derive a strong lower bound on the $k$-GM concurrence by exploiting the permutationally invariant part of a quantum state. Furthermore, we parametrize $k$-GM concurrence to obtain two more general and complete categories of quantifications, $q$-$k$-GM concurrence $(q>1)$ and $\alpha$-$k$-GM concurrence $(0\leq\alpha<1)$, which obey the properties enjoyed by $k$-GM concurrence as well. In particular, $\alpha$-$2$-GM concurrence $(0<\alpha<1)$ determines that the GHZ state and the $W$ state belong to the same hierarchy, and it is proven in detail satisfying the requirement that the GHZ state is more entangled than the $W$ state in multiqubit systems.
\end{abstract}



\maketitle

\section{Introduction}

Quantum entanglement, being a physical resource, has important consequences in quantum information processing as it enables several tasks that cannot be done within classical correlations, such as quantum computation \cite{14,15,16}, quantum cryptography \cite{7,8,9,10}, and quantum teleportation \cite{11,12,13,6}. Therefore, one of the areas of development in the theory of entanglement resource is what one characterizes quantitatively, in an unambiguous mathematical manner, the entanglement of general quantum states.

There are already some methods available for characterizing the entanglement of quantum states. For the simplest bipartite scenarios, a series of entanglement measures, such as concurrence \cite{22,23,24,25}, negativity \cite{26,27,52}, entanglement of formation \cite{28,29}, were proposed originally.
Later, Ma $et~al.$ \cite{20} introduced a genuine multipartite entanglement (GME) measure, GME concurrence, where quantification of arbitrary pure states is defined by taking the minimum of bipartite concurrence among all possible bipartite splits. Hong $et~al.$ \cite{1} put forward, whereafter, a type of $k$-nonseparable $(2\leq k\leq n)$ entanglement measures termed $k$-ME concurrence, which, for arbitrary $n$-particle pure states, by definition, are actually taking the minimum in all feasible $k$-partitions and describe more finely the whole entanglement hierarchy of $n$-partite quantum systems. More recently, two classes of parametrized $k$-nonseparable entanglement measures were defined in Ref. \cite{18}, which can be regarded as generalized forms of GME concurrence \cite{20} and $k$-ME concurrence \cite{1} in terms of pure states. In addition, there are numerous criteria to detect entanglement \cite{32,34,30,33,21,19,17,36,35,39,45}.

The structure of multipartite systems is exceedingly complicated and most of entanglement measures involve the optimization procedures for mixed states. So the computation of entanglement for $n$-partite mixed states is an enormous challenge. Fortunately, Gao $et~al$. \cite{5} claimed that $E(\rho)\geq E(\rho^{\rm PI})$ ($\rho^{\rm PI}$ denotes the permutationally invariant (PI) part of quantum state $\rho$) should be added as a requirement that any entanglement measure $E$ ought to meet and further proposed that whether a quantum state is $k$-nonseparable can be judged by its PI part, which sharply reduces the dimensionality of the considered space.

We noticed recently that a novel GME measure, geometric mean of bipartite concurrence (GBC) \cite{3}, was put forth based on the regularized fashion of bipartite concurrence \cite{31,37,38}, which owns a host of merits but for vanishing entanglement it contains all entangled states except genuinely entangled states in an indistinguishable way. Although the $k$-ME concurrence \cite{1} and parametrized $k$-ME concurrence \cite{18} are refined entanglement quantifications, they depend on the minimization procedures and neglect the global entanglement distribution among the parties, resulting that the nature of multipartite entanglement contributions cannot be characterized perfectly. Our motivation is, consequently, to present natural and powerful $k$-nonseparable quantifications in multipartite systems such that the understanding of the entanglement hierarchy is more refined and comprehensive.

The paper is structured as follows. In Sec. \ref{II}, we review some fundamental concepts. A complete type of hierarchical quantifications, called $k$-GM concurrence, are defined in Sec. \ref{III} based on geometric mean, which permit a fine graduation of states in accordance with their degrees of inseparability. Rigorous proofs show that they obey all the requisites of an entanglement measure and that GBC \cite{3} can be considered as a particular case of $k$-GM concurrence when the coefficient is not referred. In addition, comparing $k$-GM concurrence with $k$-ME concurrence, we observe that (i) $k$-GM concurrence can distinguish several entangled states in consistent hierarchy, whereas $k$-ME concurrence fails sometimes; (ii) $k$-GM concurrence is smooth; (iii) these two kinds of hierarchical measures yield different entanglement orders. Moreover, we associate $k$-GM concurrence with $k$-ME concurrence \cite{1}, and derive the result that the maximum of $k$-GM concurrence of the PI part of a quantum state is a lower bound on $k$-GM concurrence of original state. In Sec. \ref{VI}, we parametrize the $k$-GM concurrence to present two more general and full kinds of hierarchies quantifications, $q$-$k$-GM concurrence and $\alpha$-$k$-GM concurrence, in order to describe the entanglement characteristics of states from multiple dimensions. They satisfy the desired properties possessed by $k$-GM concurrence and their lower bounds are provided. In particular, we show rigorously that $\alpha$-$2$-GM concurrence $\mathcal{G}_{\alpha-2}$ $(0<\alpha<1)$ fulfills the condition that the $W$ state is less entangled than the GHZ state in multiqubit systems, and derive that the ratio of ${\mathcal{G}_{\alpha-2}(|W_n\rangle)}$ to ${{\mathcal{G}_{\alpha-2}(|{\rm GHZ}_n\rangle)}}$ tends to 1 as $n$ approaches infinity. Conclusion is summarized in Sec. \ref{VII}.

\section{Preliminaries}\label{II}
To start with, we introduce some basic notions, $k$-separable, $k$-ME concurrence, and $q$-$k$-ME concurrence, etc., which are of great necessity for subsequent sections.

The bipartite systems are the simplest ones encompassing entanglement, and $q$-concurrence is a reasonable entanglement measure associated with Tsallis-$q$ entropy $(q\geq2)$, defined as \cite{2}
\begin{equation*}
\begin{array}{rl}
C_{q}(|\psi\rangle_{AB})=1-{\rm Tr}(\rho_{A}^q)\\
\end{array}
\end{equation*}
for any bipartite pure state $|\psi\rangle_{AB}$. Whereafter, its dual quantity $\alpha$-concurrence $(0\leq\alpha\leq\frac{1}{2})$ is defined as \cite{40}
\begin{equation*}
\begin{array}{rl}
C_{\alpha}(|\psi\rangle_{AB})={\rm Tr}(\rho_{A}^\alpha)-1.\\
\end{array}
\end{equation*}
Here $\rho_A={\rm Tr}_B(|\psi\rangle_{AB}\langle\psi|)$.

The partition $A_1|A_2|\cdots|A_k$, called a $k$-partition of set $S=\{1,2,\ldots,n\}$, has to simultaneously obey the conditions: (a) the union of $A_1,A_2,\ldots,A_k$ is tantamount to the set $S$; (b) any two of them are disjoint, namely, $A_{i}\bigcap A_{j}=\emptyset~{\rm when}~i\neq j$.

An $n$-partite pure state $|\psi\rangle$ on Hilbert space $\mathcal{H}=\otimes_{i=1}^n\mathcal{H}_{i}$ (dim$\mathcal{H}_{i}=d_i$) is termed $k$-separable if it can be expressed as $|\psi\rangle=|\psi_1\rangle_{A_1}\otimes|\psi_2\rangle_{A_2}\otimes\cdots\otimes|\psi_k\rangle_{A_k}$, where $|\psi_t\rangle_{A_t}$ is a state of subsystem $A_t$ and the conditions (a) and (b) must be satisfied simultaneously among the $k$ subsystems. An $n$-partite mixed state $\rho$ is known $k$-separable if it can be represented as a convex mixture of $k$-separable pure states, i.e., $\rho=\sum_{i}p_i|\psi_i\rangle\langle\psi_i|$ with $|\psi_i\rangle$ possibly being $k$-separable respecting different partitions. Or else the quantum state $\rho$ is called $k$-nonseparable. In particular, if $\rho$ is 2-nonseparable, then it is called genuinely entangled state.

Let $S_k$ be the set of $k$-separable $(k=2,3,\ldots,n)$ quantum states and $S_1$ be the set containing all quantum states, the relation between them is $S_n\subset S_{n-1}\subset\cdots\subset S_2\subset S_1$, $S_1\backslash S_k$ stands for the set consisting of all $k$-nonseparable states.

For any $n$-partite pure state $|\psi\rangle$, the $k$-ME concurrence is defined as \cite{1}
\begin{equation}
\begin{array}{rl}
C_{k-{\rm ME}}(|\psi\rangle)=\min\limits_A\sqrt{\frac{2\sum_{t=1}^{k}(1-{\rm Tr}\rho_{A_t}^2)}{k}},\\
\end{array}
\end{equation}
and the $q$-$k$-ME concurrence ($q>1$) is defined as \cite{18}
\begin{equation}\label{21}
\begin{array}{rl}
C_{q-k}(|\psi\rangle)=\min\limits_A\frac{\sum_{t=1}^{k}[1-{\rm Tr}(\rho_{A_t}^q)]}{k}.
\end{array}
\end{equation}
Here $\rho_{A_t}={\rm Tr}_{{\overline A}_t}(|\psi\rangle\langle\psi|)$, ${\overline A}_t$ is the complement of $A_t$, and the minimum runs over all feasible $k$-partitions $A=\{A_1|A_2|\cdots|A_k\}$. Let $\mathcal{P}_k(|\psi\rangle)=\sqrt{\frac{{2\sum_{t=1}^{k}(1-{\rm Tr}\rho_{A_t}^2)}}{k}}$, we refer to this as the entanglement value of $k$-partition.

For any $n$-partite mixed state $\rho$, the $k$-ME concurrence is defined, as shown in Ref. \cite{1}, by convex-roof extension
\begin{equation}
\begin{array}{rl}
C_{k-{\rm ME}}(\rho)=\inf\limits_{\{p_i, |\psi_i\rangle\}}\sum_ip_iC_{k-\rm ME}(|\psi_i\rangle),\\
\end{array}
\end{equation}
and the $q$-$k$-ME concurrence is given as \cite{18}
\begin{equation}\label{13}
\begin{array}{rl}
C_{q-k}(\rho)=\inf\limits_{\{p_i, |\psi_i\rangle\}}\sum_ip_iC_{q-k}(|\psi_i\rangle).\\
\end{array}
\end{equation}
Here the infimum is done over all possible pure state decompositions.

Let $E$ be an entanglement measure and $\rho$ be any quantum state, if there is a decomposition $\{p_i,\rho_i\}$ of $\rho$ such that $E(\rho)=\sum_ip_iE(\rho_i)$, then $\{p_i,\rho_i\}$ is called the optimal decomposition of $E(\rho)$.

\section{A whole type of hierarchical entanglement measures}\label{III}
For bipartite pure state $|\psi\rangle_{AB}$, concurrence, a widely used entanglement measure, is defined as \cite{23}
\begin{equation*}
\begin{array}{rl}
C(|\psi\rangle_{AB})=\sqrt{2(1-{\rm Tr\rho_A^2})},\\
\end{array}
\end{equation*}
where $\rho_A={\rm Tr}_B(|\psi\rangle_{AB}\langle\psi|)$. Moreover, concurrence is essential in remote entanglement distribution protocols, such as entanglement swapping and remote preparation of bipartite entangled states, shown by the 2004 work of Gour and Sanders \cite{49}. Here we will attempt to generalize concurrence to provide a full set of hierarchical multipartite entanglement quantifications as a way to finely characterize graduation of states in accordance with entanglement contributions among parties.
\subsection{$k$-GM concurrence}
{\bf Definition 1}. For any $n$-partite pure state $|\psi\rangle$, the $k$-GM concurrence $(2\leq k\leq n)$ is defined as
\begin{equation}\label{1}
\begin{array}{rl}
\mathcal{G}_{k-{\rm GM}}(|\psi\rangle)&=\big[\prod_{\alpha_i\in T_k}\mathcal{P}_k(|\psi\rangle)\big]^{1/|T_k|},\\
\end{array}
\end{equation}
where
\begin{equation*}
\begin{array}{rl}
\mathcal{P}_k(|\psi\rangle)=\sqrt{\frac{2\sum_{t=1}^kC_{2A_{t{\alpha_i}}|\overline{A}_{t{\alpha_i}}}(|\psi\rangle)}{k}}
\end{array}
\end{equation*}
and
\begin{equation*}
\begin{array}{rl}
|T_k|=\sum_{t=1}^k\frac{(-1)^{k-t}t^{n-1}}{(t-1)!(k-t)!},
\end{array}
\end{equation*}
$T_k=\{\alpha_i\}$ is the set that stands for all possible $k$-partitions $\{A_{1\alpha_i}|A_{2\alpha_i}|\cdots|A_{k\alpha_i}\}$, $|T_k|$ is the Stirling number of the second kind \cite{51} used to represent the cardinality of the elements in the set $T_k$, $C_2$ is the special case of $q$-concurrence corresponding to $q=2$, ${A_{t{\alpha_i}}}|{{\overline A}_{t{\alpha_i}}}$ denotes any split of state $|\psi\rangle$, ${\overline A}_{t{\alpha_i}}$ is the complement of $A_{t{\alpha_i}}$.

In Table \ref{tab:test}, we present the cardinalities of $k$-partition $(2\leq k\leq n)$ for $n$-partite $(n=5,6,7,8)$ quantum systems.

In contrast to $k$-ME concurrence \cite{1}, the hierarchical entanglement quantifications given in Definition 1 do not rely on the processes of minimization, so they can be said to be a more comprehensive description of the entanglement characteristics of multipartite quantum states.

Eq. (\ref{1}) can also be directly denoted as
\begin{equation*}
\begin{array}{rl}
\mathcal{G}_{k-{\rm GM}}(|\psi\rangle)=\frac{\big({\prod_{\alpha_i\in T_k}[{2\sum_{t=1}^kC_{2{{A_{t{\alpha_i}}}|{{\overline A}_{t{\alpha_i}}}}}(|\psi\rangle)}]^{1/2}}\big)^{1/|T_k|}}{\sqrt{k}}.\\
\end{array}
\end{equation*}
When $k=2$, $2$-GM concurrence can be reduced as
\begin{equation}\label{2}
\begin{array}{rl}
\mathcal{G}_{2-{\rm GM}}(|\psi\rangle)&=\big\{\prod_{\alpha_i\in T_2}[2C_{2{{A_{t{\alpha_i}}}|{{\overline A}_{t{\alpha_i}}}}}(|\psi\rangle)]^{1/2}\big\}^{1/|T_2|}\\
&=\big[\prod_{\alpha_i\in T_2}C_{{{A_{t{\alpha_i}}}|{{\overline A}_{t{\alpha_i}}}}}(|\psi\rangle)\big]^{1/|T_2|},\\
\end{array}
\end{equation}
where
\begin{equation*}
|T_2|=\begin{cases} \sum_{p=1}^{{(n-1)}/{2}}C_{n}^p, &n~{\rm is~odd},\\
\sum_{p=1}^{{(n-2)}/{2}}C_{n}^p+\frac{1}{2}C_n^{\frac{n}{2}}, &n{\rm~is~even},\end{cases}
\end{equation*}
and $C_{n}^p=\frac{n(n-1)\cdots(n-p+1)}{p(p-1)\cdots1}$. In fact, $|T_2|=2^{n-1}-1$.

For any 3-qubit pure state $|\psi\rangle$, the $2$-GM concurrence $\mathcal{G}_{2-{\rm GM}}(|\psi\rangle)$ is accorded with the geometric mean of bipartite concurrence (GBC) $\mathcal{G}(|\psi\rangle)$ defined in Ref. \cite{3}, i.e., $\mathcal{G}_{2-{\rm GM}}(|\psi\rangle)=\mathcal{G}(|\psi\rangle)$. However, for the rest of cases, 3-partite systems $\mathcal{H}_{1}\otimes\mathcal{H}_{2}\otimes\mathcal{H}_{3}$ $({\rm dim}\mathcal{H}_{i}>2)$ or $n$-partite systems $\otimes_{i=1}^n\mathcal{H}_{i}$ ($n>3,~{\rm dim}\mathcal{H}_{i}\geq2$), due to the intricacy of the partition of multipartite quantum systems and the dependence of GBC on the smallest dimension of subsystem, the relation between $\mathcal{G}$ and $\mathcal{G}_{2-{\rm GM}}$ becomes complicated for any $n$-partite pure state $|\varphi\rangle$, which is
\begin{table}[htbp]
\centering
 \caption{\label{tab:test} The cardinalities of all possible $k$ partitions $(2\leq k\leq n)$ are listed for $n$-partite quantum systems $(n=5,6,7,8)$.}
 \begin{tabular}{cccccccc}
  \hline\hline

 ~~&~~~$\emph{k}$=2~~  & $~~\emph{k}$=3~~ & $~~\emph{k}$=4~~  & $~~\emph{k}$=5~~ & $~~\emph{k}$=6~~ & $~~\emph{k}$=7~~ & $\emph{k}$=8\\
 \hline
 \vspace{0.5em}
 $n=5$ & 15 & 25 & 10 & 1 & - & - & -\\
 \vspace{0.5em}
 $n=6$ & 31 & 90 & 65 & 15 & 1 & - & -\\
 \vspace{0.5em}
 $n=7$ & 63 & 301 & 350 & 140 & 21 & 1 & -\\
 \vspace{0.5em}
 $n=8$ & 127 & 966 & 1701 & 1050 & 266 & 28 & 1\\
  \hline\hline
 \end{tabular}
\end{table}
\begin{equation*}
\begin{array}{rl}
\mathcal{G}_{2-{\rm GM}}(|\varphi\rangle)=\Big({\prod_{i=1}^{|T_2|}\frac{2(D_i-1)}{D_i}}\Big)^{1/|T_2|}\mathcal{G}(|\varphi\rangle),\\
\end{array}
\end{equation*}
where $D_i$ denotes the minimum dimension of subsystem for each partition, i.e., $D_i=\min\{{\rm dim}{A_{t{\alpha_i}}}, {\rm dim}\overline{A}_{t{\alpha_i}}\}$. This means that GBC \cite{3} can be considered as equivalent to 2-GM concurrence. Therefore, the genuine entanglement measure, GBC \cite{3}, is a special case of $k$-GM concurrence. The $2$-GM concurrence can be reduced to concurrence \cite{23} when $n=2$.

The $k$-GM concurrence can be generalized to $n$-partite mixed state by convex-roof extension
\begin{equation}\label{3}
\begin{array}{rl}
\mathcal{G}_{k-{\rm GM}}(\rho)=\inf\limits_{\{p_i,|\psi_i\rangle\}}\sum_ip_i\mathcal{G}_{k-{\rm GM}}(|\psi_i\rangle),\\
\end{array}
\end{equation}
where the infimum is done over all possible pure state decompositions.

Note that the $n$-GM concurrence is the same as $n$-ME concurrence \cite{1} due to $|T_n|=1$.

Next we demonstrate that $k$-GM concurrence satisfy the necessary conditions to be an entanglement measure.

{\bf Theorem 1}. The $k$-GM concurrence $\mathcal{G}_{k-{\rm GM}}$ is a legitimate type of hierarchical entanglement measures satisfying the properties: (M1) $\mathcal{G}_{k-{\rm GM}}(\rho)=0$ for any $\rho\in S_k$; (M2) $\mathcal{G}_{k-{\rm GM}}(\rho)>0$ for any $\rho\in S_1\backslash S_k$; (M3) $\mathcal{G}_{k-{\rm GM}}(\rho)$ is invariant under local unitary transformation; (M4) $\mathcal{G}_{k-{\rm GM}}(\rho)$ is non-increasing under LOCC (called entanglement monotone); (M5) $\mathcal{G}_{k-{\rm GM}}(\rho)$ does not increase on average under LOCC (termed strong monotone); (M6) $\mathcal{G}_{k-{\rm GM}}(\rho)$ is convex with respect to $\rho$.

{\bf Proof}. It is easy to show that the properties (M3) and (M6) hold, and we will elaborate that $k$-GM concurrence satisfies the rest of properties.

(M1) Suppose that $|\psi\rangle$ is an arbitrary pure state belonging to $S_k$, there must exist some $k$-partition such that $\sum_{t=1}^kC_{2{{A_{t{\alpha_i}}}|{{\overline A}_{t{\alpha_i}}}}}(|\psi\rangle)=0$, then $\mathcal{G}_{k-{\rm GM}}(|\psi\rangle)=0$. Let $\rho$ be an arbitrary $k$-separable mixed state and $\{p_i,|\psi_i\rangle\}$ be any pure state decomposition of $\rho$, where $|\psi_i\rangle\in S_k$, one has $\mathcal{G}_{k-{\rm GM}}(\rho)\leq\sum_i p_i\mathcal{G}_{k-{\rm GM}}(|\psi_i\rangle)=0$. Based on the above analysis, we can see that $\mathcal{G}_{k-{\rm GM}}(\rho)=0$ for arbitrary $\rho\in S_k$.

(M2) Let $|\psi\rangle$ be any $k$-nonseparable pure state, then there is always $\sum_{t=1}^kC_{2{{A_{t{\alpha_i}}}|{{\overline A}_{t{\alpha_i}}}}}(|\psi\rangle)>0$ under any $k$-partition, so $\mathcal{G}_{k-{\rm GM}}(|\psi\rangle)>0$. For any mixed state $\rho\in S_1\backslash S_k$, it cannot be written as a convex mixture of $k$-separable pure states, hence $\mathcal{G}_{k-{\rm GM}}(\rho)>0$ holds evidently.

(M4) We first show that the relation $\mathcal{G}_{k-{\rm GM}}(|\psi\rangle)\geq\mathcal{G}_{k-{\rm GM}}(\Lambda_{\rm LOCC}(|\psi\rangle))$ holds for any LOCC acting on pure state $|\psi\rangle$. Since $q$-concurrence is non-increasing under LOCC \cite{2}, we just require to verify that $\mathcal{G}_{k-{\rm GM}}(|\psi\rangle)$ is an increasing function of $C_{2A_{t{\alpha_i}}|\overline{A}_{t{\alpha_i}}}(|\psi\rangle)$. Through calculation, the following result can be obtained,
\begin{equation*}
\begin{array}{rl}
&\frac{\partial\mathcal{G}_{k-{\rm GM}}(|\psi\rangle)}{\partial C_{2A_{t\alpha_l}|{\overline A}_{t\alpha_l}}(|\psi\rangle)}\\
=&\frac{\prod_{T_k\backslash \{\alpha_l\}}[{\sum_{t=1}^kC_{2{{A_{t{\alpha_i}}}|{{\overline A}_{t{\alpha_i}}}}}(|\psi\rangle)}]}{\sqrt{2k}|T_k|\big(\prod_{\alpha_i\in T_k}[{\sum_{t=1}^kC_{2{{A_{t{\alpha_i}}}|{{\overline A}_{t{\alpha_i}}}}}(|\psi\rangle)}]\big)^{\frac{2|T_k|-1}{2|T_k|}}}\\
\geq&0,
\end{array}
\end{equation*}
where $l=1,2,\ldots,|T_k|$. This implies that the monotonicity of $\mathcal{G}_{k-{\rm GM}}(|\psi\rangle)$ is true, so $\mathcal{G}_{k-{\rm GM}}(|\psi\rangle)$ is non-increasing under LOCC.

Let $\rho$ be an arbitrary mixed state with $\{p_i,|\psi_i\rangle\}$ being the optimal pure decomposition of $\mathcal{G}_{k-{\rm GM}}(\rho)$, then one obtains
\begin{equation*}
\begin{array}{rl}
&\mathcal{G}_{k-{\rm GM}}(\Lambda_{\rm LOCC}(\rho))\\
=&\mathcal{G}_{k-{\rm GM}}(\Lambda_{\rm LOCC}(\sum_{i}p_i|\psi_i\rangle\langle\psi_i|))\\
=&\mathcal{G}_{k-{\rm GM}}(\sum_{i}p_i\Lambda_{\rm LOCC}(|\psi_i\rangle\langle\psi_i|))\\
\leq&\sum_{i}p_i\mathcal{G}_{k-{\rm GM}}(\Lambda_{\rm LOCC}(|\psi_i\rangle))\\
\leq&\sum_{i}p_i\mathcal{G}_{k-{\rm GM}}(|\psi_i\rangle)\\
=&\mathcal{G}_{k-{\rm GM}}(\rho).
\end{array}
\end{equation*}
Here the first inequality is owing to the convexity of $k$-GM concurrence, the second inequality holds according to the property that $k$-GM concurrence does not increase under LOCC for any pure state.

Therefore, $\mathcal{G}_{k-{\rm GM}}(\rho)$ complies with the monotonicity.

(M5) Since $q$-concurrence satisfies strong monotonicity \cite{2}, that is, $C_q(\rho)\geq \sum_j p_j C_q(\sigma_{j})$, where the state $\sigma_j$ is produced with the probability $p_j$ via performing LOCC on $\rho$. If $\rho=|\psi\rangle\langle\psi|$ is a pure state, then $\sigma_j=K_j|\psi\rangle\langle\psi|K_j^\dagger$ is also a pure state, $\sum_jK_j^\dagger K_j=\mathbb{I}$ (unit operator), and one gets
\begin{equation*}
\begin{array}{rl}
\mathcal{G}_{k-{\rm GM}}(\rho)&=\frac{\big({\prod_{{\alpha_i}\in T_k}[2\sum_{t=1}^kC_{2A_{t{\alpha_i}}|{\overline A}_{t{\alpha_i}}}(|\psi\rangle)]^{1/2}}\big)^{1/|T_k|}}{\sqrt{k}}\\
&\geq\frac{\big({\prod_{{\alpha_i}\in T_k}[2\sum_{t=1}^k\sum_{j}p_{j}C_{2A_{t{\alpha_i}}|{\overline A}_{t{\alpha_i}}}(\sigma_{j})]^{1/2}}\big)^{1/|T_k|}}{\sqrt{k}}\\
&=\frac{\big(\prod_{\alpha_i\in T_k}\{\sum_jp_j[2\sum_{t=1}^kC_{2{A_{t{\alpha_i}}|{\overline A}_{t{\alpha_i}}}}(\sigma_{j})]\}^{1/2}\big)^{1/|T_k|}}{\sqrt{k}}\\
&\geq\frac{\big(\prod_{\alpha_i\in T_k}\{\sum_jp_j[2\sum_{t=1}^kC_{2{A_{t{\alpha_i}}|{\overline A}_{t{\alpha_i}}}}(\sigma_{j})]^{1/2}\}\big)^{1/|T_k|}}{\sqrt{k}}\\
&\geq\frac{\sum_jp_j\big(\prod_{\alpha_i\in T_k}[2\sum_{t=1}^kC_{2{A_{t{\alpha_i}}|{\overline A}_{t{\alpha_i}}}}(\sigma_{j})]^{1/2}\big)^{1/|T_k|}}{\sqrt{k}}\\
&\geq\sum_{j}p_{j}\mathcal{G}_{k-{\rm GM}}(\sigma_{j}).\\
\end{array}
\end{equation*}
Here the second inequality follows from the concavity of the function $y=x^{1/2}$, the third inequality is true because the geometric mean function $f=\big({\prod_{i=1}^{n}x_i}\big)^{1/n}$ is a concave function which can be proved by verifying that the Hessian matrix formed by the second partial derivatives of $-f$ is semi-positive definite \cite{4}, and the last inequality is clearly true by definition.

Assume that $\{p_i,|\psi_i\rangle\}$ is the optimal pure decomposition of $\mathcal{G}_{k-{\rm GM}}(\rho)$ and the LOCC is given by Kraus operators $K_j$, $\sum_jK_j^\dag K_j=\mathbb{I}$,  then we have
\begin{equation*}
\begin{array}{rl}
&\mathcal{G}_{k-{\rm GM}}(\rho)=\sum_ip_i\mathcal{G}_{k-{\rm GM}}(|\psi_{i}\rangle)\\
\geq&\sum_{ij}p_i{\rm Tr}(K_j|\psi_i\rangle\langle\psi_i|K_j^\dag)\mathcal{G}_{k-{\rm GM}}\Big(\frac{K_j|\psi_i\rangle\langle\psi_i|K_j^\dag}{{\rm Tr}(K_j|\psi_i\rangle\langle\psi_i|K_j^\dag)}\Big)\\
=&\sum_{ij}{{\rm Tr}(K_j\rho K_j^\dag)}\frac{p_i{{\rm Tr}(K_j|\psi_i\rangle\langle\psi_i|K_j^\dag)}}{{{\rm Tr}(K_j\rho K_j^\dag)}}\\
&\times\mathcal{G}_{k-{\rm GM}}\Big(\frac{K_j|\psi_i\rangle\langle\psi_i|K_j^\dag}{{\rm Tr}(K_j|\psi_i\rangle\langle\psi_i|K_j^\dag)}\Big)\\
=&\sum_{j}p_j[\sum_ip_{ij}\mathcal{G}_{k-{\rm GM}}(|\psi_{ij}\rangle)]\\
\geq&\sum_jp_j\mathcal{G}_{k-{\rm GM}}(\sigma_j),\\
\end{array}
\end{equation*}
where $|\psi_{ij}\rangle=\frac{K_j|\psi_i\rangle}{{\sqrt {{\rm Tr}(K_j|\psi_i\rangle\langle\psi_i|K_j^\dag)}}}$,~$p_{ij}=\frac{p_i{{\rm Tr}(K_j|\psi_i\rangle\langle\psi_i|K_j^\dag)}}{{{\rm Tr}(K_j\rho K_j^\dag)}}$, and the state $\sigma_j=\sum_ip_{ij}|\psi_{ij}\rangle\langle\psi_{ij}|$ occurs with the probability $p_{j}={\rm Tr}(K_j\rho K_j^\dag)$ by LOCC acting on $\rho$. The first inequality holds since $k$-GM concurrence obeys the strong monotonicity for any pure state, the second inequality is due to the definition of $\mathcal{G}_{k-{\rm GM}}(\rho)$. $\hfill\blacksquare$

Based on the analysis above, we know that $k$-GM concurrence, as a set of indicators, can not only classify $n$-partite quantum states into fully separable states and entangled states, but also classify $n$-partite entangled states into 2-nonseparable states (genuine entangled states), 3-nonseparable states, $\ldots$ , $n$-nonseparable states, and $k$-GM concurrence enjoys all the requirements of an entanglement measure. These properties are conducive to making it a potential quantum resource, and its simplicity may be more commodious for practical applications.

A question may be raised as to how $k$-GM concurrence differs from $k$-ME concurrence, which is analyzed in the following subsection

\subsection{The comparison between $k$-ME concurrence and $k$-GM concurrence}\label{IV}
Compared to $k$-ME concurrence \cite{1}, $k$-GM concurrence manifests some differences, which are elaborated from three aspects.

First, there are some quantum states that can be distinguished by $k$-GM concurrence but contain the same amount of entanglement when evaluated utilizing $k$-ME concurrence. For example, consider the quantum states $|\psi_1\rangle=\frac{1}{2}(|0000\rangle+|1011\rangle+|1101\rangle+|1111\rangle)$ and $|\psi_2\rangle=\frac{1}{2}(|0000\rangle+|1001\rangle+|1110\rangle+|1111\rangle)$, by calculation, we can get $C_{2-\rm ME}(|\psi_1\rangle)=C_{2-\rm ME}(|\psi_2\rangle)=\frac{\sqrt3}{2}$, $C_{3-\rm ME}(|\psi_1\rangle)=C_{3-\rm ME}(|\psi_2\rangle)=\frac{\sqrt3}{2}$, while $\mathcal{G}_{2-{\rm GM}}(|\psi_1\rangle)=\frac{\sqrt[14]{3888}}{2}\neq\mathcal{G}_{2-{\rm GM}}(|\psi_2\rangle)=\frac{\sqrt[14]{10800}}{2}$, $\mathcal{G}_{3-{\rm GM}}(|\psi_1\rangle)=\frac{\sqrt[12]{590490000}}{6}\neq\mathcal{G}_{3-{\rm GM}}(|\psi_2\rangle)=\frac{\sqrt[12]{2816}}{2}$.

\begin{figure}[htbp]
\centering
{\includegraphics[width=8.5cm,height=6.1cm]{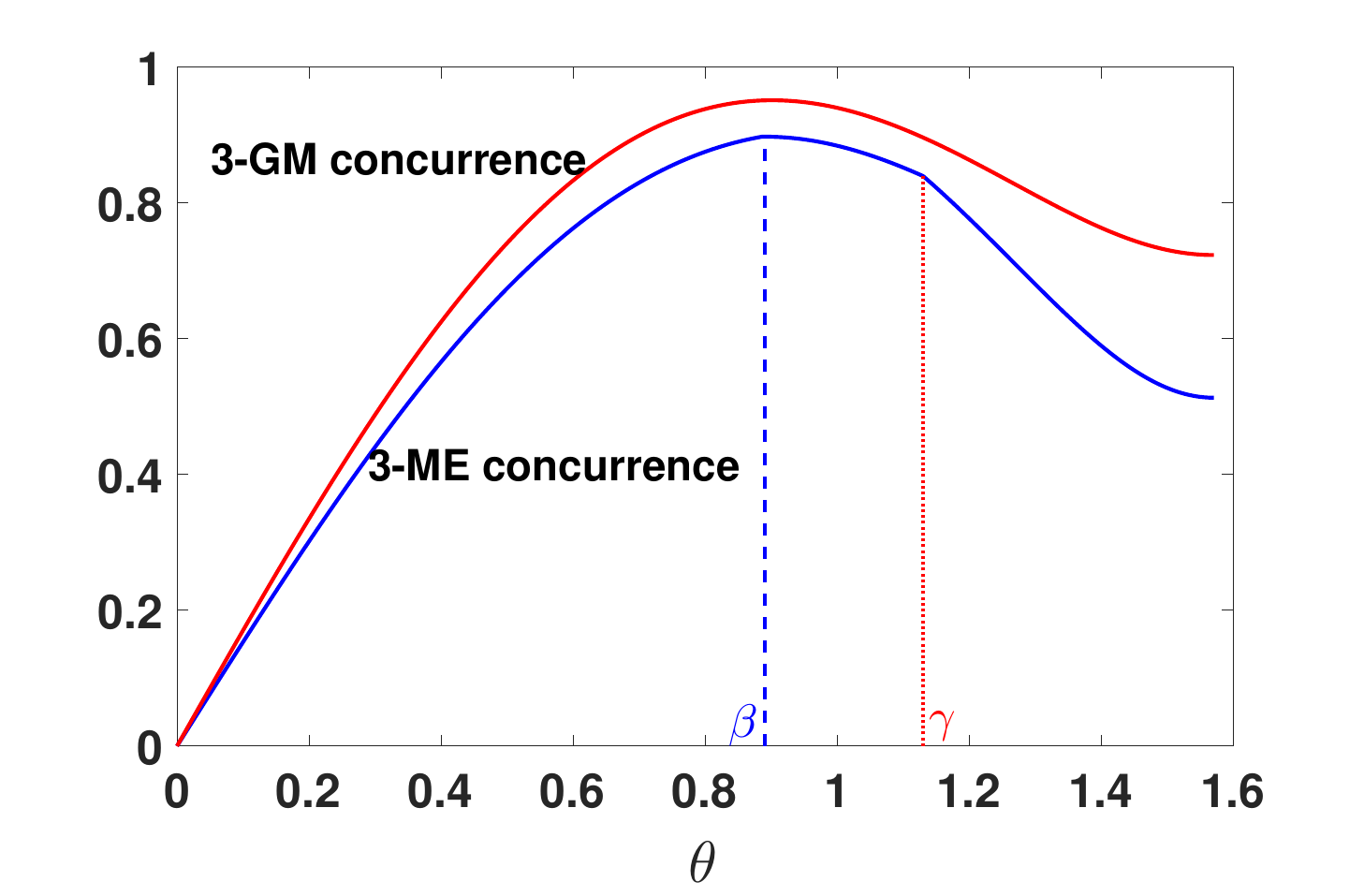}}
\caption{The red (upper) curve represents $3$-GM concurrence with respect to $|\psi\rangle$, the blue (lower) curve denotes $3$-ME concurrence in term of $|\psi\rangle$. When $\theta=\beta$ or $\gamma$, we find the curve of $3$-ME concurrence is not smooth.} \label{fig 1}
\end{figure}
Second, the $k$-GM concurrence does not involve the minimization processes in terms of pure states, which leads to that $k$-GM concurrence may be computationally simpler than $k$-ME concurrence  and that $k$-GM concurrence is smooth. The smoothness means that the measure does not produce sharp peaks if the pure state being measured varies continuously. For instance, considering a 4-qubit pure state $|\psi\rangle={\rm sin}\theta\big(\frac{1}{3}|0001\rangle+\frac{\sqrt2}{3}|0100\rangle+\frac{\sqrt6}{3}|1000\rangle\big)+{\rm cos}\theta|0011\rangle$, we observe, as shown in Fig. \ref{fig 1}, $k$-GM concurrence is smooth, while $k$-ME concurrence emerges two discontinuity points.

\begin{figure}[htbp]
\centering
{\includegraphics[width=8.5cm,height=6.1cm]{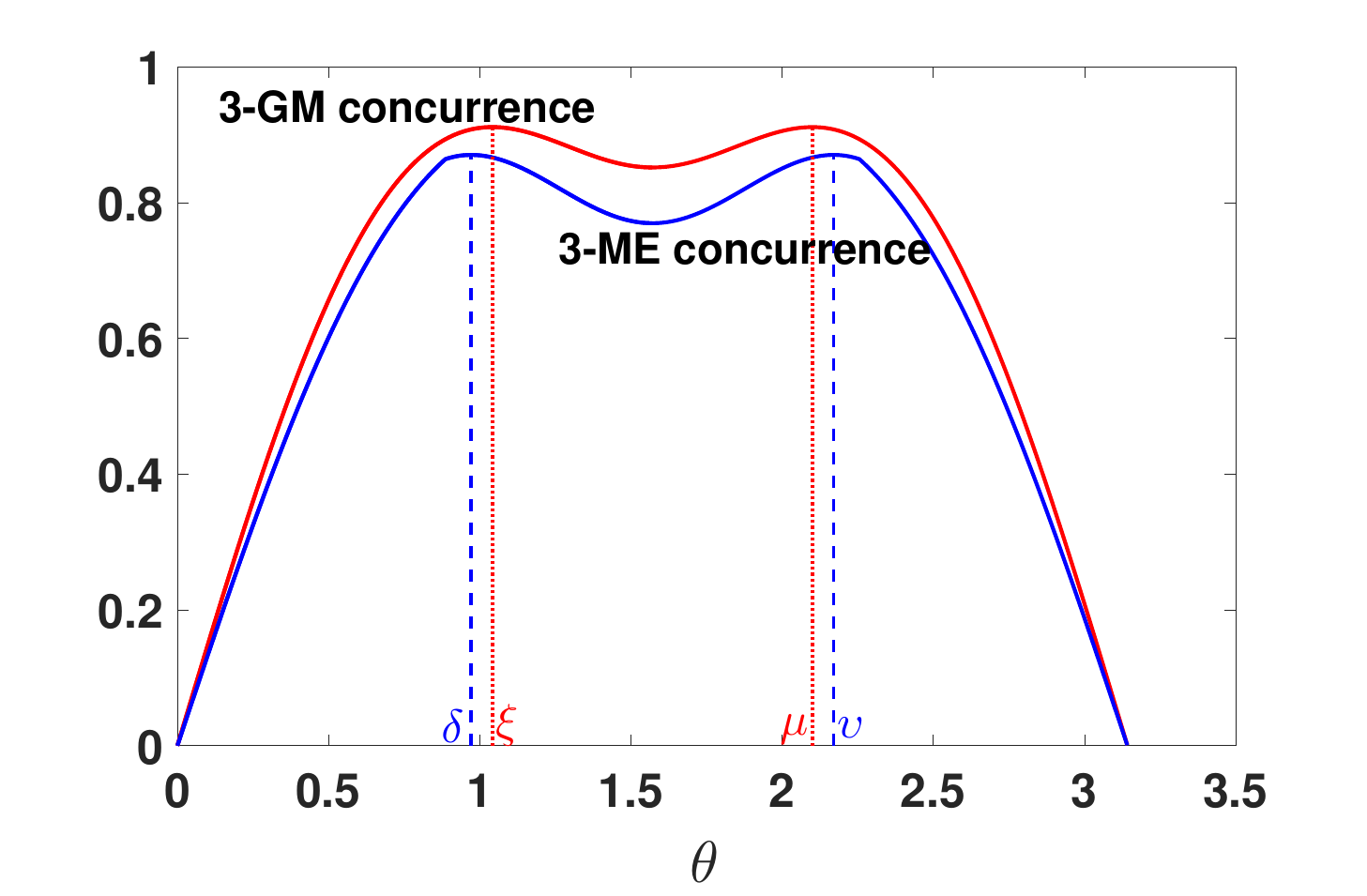}}
\caption{The red (upper) curve stands for $3$-GM concurrence with respect to $|\phi_\theta\rangle$, the blue (lower) curve is $3$-ME concurrence in terms of $|\phi_\theta\rangle$. When $\theta=\theta_1,\theta_2$ are selected from either $(\delta,\xi)$ or $(\mu,\upsilon)$, the entanglement orders of $\mathcal{G}_{k-{\rm GM}}(|\phi_\theta\rangle)$ and $C_{k-\rm ME}(|\phi_\theta\rangle)$ are different. And the blue line exist two points which are not smooth.\\}\label{fig 2}
\end{figure}

Third, these two kinds of entanglement measures may yield different entanglement orders, that is, there exist $|\phi_1\rangle$ and $|\phi_2\rangle$ such that $C_{k-\rm ME}(|\phi_1\rangle)\geq C_{k-\rm ME}(|\phi_2\rangle)$, while $\mathcal{G}_{k-{\rm GM}}(|\phi_1\rangle)\leq \mathcal{G}_{k-{\rm GM}}(|\phi_2\rangle)$. In order to make it more intuitive, we will illustrate the fact with a concrete example. Given a 4-qubit pure state $|\phi_\theta\rangle=\frac{\sqrt3}{3}{\rm sin}\theta(|0001\rangle+|0100\rangle+|1000\rangle)+{\rm cos}\theta|0011\rangle$, computing $3$-GM concurrence and $3$-ME concurrence with respect to $|\phi_\theta\rangle$ and plotting them in Fig. \ref{fig 2}, we can see that $\mathcal{G}_{k-{\rm GM}}(|\phi_\theta\rangle)$ and $C_{k-\rm ME}(|\phi_\theta\rangle)$ have distinct entanglement orders when $\theta$ is taken as $\theta_1$ and $\theta_2$ belonging to either $(\delta,\xi)$ or $(\mu,\upsilon)$.

Consequently, from the first and second points stated above, the $k$-GM concurrence does exhibit advantages over the $k$-ME concurrence at times.

However, the structure of the multipartite quantum states is immensely complicated and $k$-GM concurrence involves the procedures of taking the infimum for mixed states, resulting the computation of $k$-GM concurrence for $n$-partite mixed states as a huge challenge. We strive to give its lower bounds in next subsection.
\subsection{The lower bounds of $k$-GM concurrence}\label{V}
In Ref. \cite{1}, a category of multipartite entanglement measures, $k$-ME concurrence $C_{k-{\rm ME}}$ $(2\leq k\leq n)$, were advanced. We observe that a link between $k$-GM concurrence and $k$-ME concurrence can be established, as shown in the following theorem.

{\bf Theorem 2}. The $k$-ME concurrence is a lower bound of the $k$-GM concurrence.

{\bf Proof}. Let $|\psi\rangle$ be any pure state of $n$-particle, by the definition of $k$-GM concurrence and $k$-ME concurrence, we can directly obtain the relation between them as
\begin{equation*}
\begin{array}{rl}
\mathcal{G}_{k-{\rm GM}}(|\psi\rangle)\geq C_{k-{\rm ME}}(|\psi\rangle).
\end{array}
\end{equation*}

If $\rho$ is an arbitrary mixed state and $\{p_j,|\psi_j\rangle\}$ is the optimal pure decomposition of $\mathcal{G}_{k-{\rm GM}}(\rho)$, then one has
\begin{equation*}
\begin{array}{rl}
\mathcal{G}_{k-{\rm GM}}(\rho)=&\sum_jp_j\mathcal{G}_{k-{\rm GM}}(|\psi_j\rangle)\\
\geq&\sum_jp_jC_{k-{\rm ME}}(|\psi_j\rangle)\\
\geq&C_{k-{\rm ME}}(\rho),\\
\end{array}
\end{equation*}
where the last inequality is due to the definition of $C_{k-{\rm ME}}(\rho)$ shown in Eq. (\ref{13}). $\hfill\blacksquare$

Gao $et~al$. \cite{5} argued that $E(\rho)\geq E(\rho^{\rm PI})$ should be added as a condition which any multipartite entanglement measure $E$ ought to conform, but this lower bound is weak since $E(\rho^{\rm PI})$ may be equal zero even if $\rho$ is maximally entangled state. Thereupon, Gao $et~al$. \cite{5} further improved it and obtained that $k$-ME concurrence of $\rho$ is greater than or equal to the maximum of $k$-ME concurrence of the PI part of $\rho$, which is a stronger lower bound. However, for any individual presented entanglement measure one must testify whether it does. Next, we discuss whether the $k$-GM concurrence meets this requirement.

{\bf Theorem 3}. The $k$-GM concurrence of an $n$-particle quantum state $\rho$ is lower bounded by the the maximum of $k$-GM concurrence of $\rho_U^{\rm PI}$, namely,
\begin{equation*}
\begin{array}{rl}
\mathcal{G}_{k-{\rm GM}}(\rho)\geq\max\limits_{U}\mathcal{G}_{k-{\rm GM}}(\rho_U^{\rm PI}),
\end{array}
\end{equation*}
where $\rho_U^{\rm PI}=(U\rho U^\dagger)^{\rm PI}$ and $U$ is any locally unitary transformation.

{\bf Proof}. If the set $\{1,2,\ldots,n\}$ is divided into $k$ parts, i.e., $A_1|A_2|\cdots|A_k$, which satisfy conditions (a) and (b), then $\Pi_{j}(A_1)|\Pi_{j}(A_2)|\cdots|\Pi_{j}(A_k)$, meanwhile, is a $k$-partition of $\{1,2,\ldots,n\}$. Let $|\psi\rangle$ be an arbitrary pure state, then $\Pi_j(|\psi\rangle)$ is a pure state as well, and one has
\begin{equation}\label{6}
\begin{array}{rl}
\mathcal{G}_{k-{\rm GM}}(|\psi\rangle)=\mathcal{G}_{k-{\rm GM}}(\Pi_j(|\psi\rangle)).
\end{array}
\end{equation}
Here $\Pi_j$ is an arbitrary element of $n$-order symmetric group.

Since $k$-GM concurrence possesses convexity and obeys Eq. (\ref{6}), we obtain
\begin{equation}\label{7}
\begin{array}{rl}
\mathcal{G}_{k-{\rm GM}}(\rho^{\rm PI})&\leq \frac{1}{n!}\sum_{j=1}^{n!}\mathcal{G}_{k-{\rm GM}}(\Pi_j|\psi\rangle)\\
&=\frac{1}{n!}\sum_{j=1}^{n!}\mathcal{G}_{k-{\rm GM}}(|\psi\rangle)\\
&=\mathcal{G}_{k-{\rm GM}}(|\psi\rangle).
\end{array}
\end{equation}
Here $\rho^{\rm PI}=\frac{1}{n!}\sum_{j=1}^{n!}\Pi_j|\psi\rangle\langle\psi|\Pi_j^\dagger$.

Let $\rho$ be an arbitrary mixed state with $\{p_i,|\psi_i\rangle\}$ being the optimal pure state decomposition of $\mathcal{G}_{k-{\rm GM}}(\rho)$, and $\rho^{\rm PI}=\sum_ip_i(|\psi_i\rangle\langle\psi_i|)^{\rm PI}$ \cite{5}, then one has
\begin{equation}\label{8}
\begin{array}{rl}
\mathcal{G}_{k-{\rm GM}}(\rho)&=\sum_{i}p_{i}\mathcal{G}_{k-{\rm GM}}(|\psi_{i}\rangle)\\
&\geq\sum_{i}p_{i}\mathcal{G}_{k-{\rm GM}}(\rho_{i}^{\rm PI})\\
&\geq \mathcal{G}_{k-{\rm GM}}(\rho^{\rm PI}),\\
\end{array}
\end{equation}
where the first inequality is following from inequality (\ref{7}) and the second inequality holds owing to the convexity of $\mathcal{G}_{k-{\rm GM}}(\rho)$.

Since the PI part relies on the choice of basis, Eq. (\ref{6}) and inequalities (\ref{7}), (\ref{8}) are valid for arbitrary locally unitary transformations, one sees
\begin{equation}
\begin{array}{rl}
\mathcal{G}_{k-{\rm GM}}(\rho)\geq\max\limits_U \mathcal{G}_{k-{\rm GM}}(\rho_U^{\rm PI}).\\
\end{array}
\end{equation}

The proof is completed. $\hfill\blacksquare$

Accordingly, the $k$-GM concurrence fulfills the requirement set forth in Ref. \cite{5}.

\section{Parametrized $k$-GM concurrence}\label{VI}
The research on parametrized entanglement measures has attracted widespread attention more recently, such as $q$-concurrence ($q\geq2$) \cite{2}, $\alpha$-concurrence ($0\leq\alpha\leq\frac{1}{2}$) \cite{40}, $\alpha$-logarithmic negativity ($\alpha\geq1$) \cite{50}, etc. With this in mind, it seems fruitful to generalize $k$-GM concurrence $(2\leq k\leq n)$ aforementioned to obtain two families of parametrized hierarchical entanglement measures called $q$-$k$-GM concurrence $(q>1,2\leq k\leq n)$ and $\alpha$-$k$-GM concurrence $(0\leq\alpha<1,2\leq k\leq n)$, respectively, for $n$-partite quantum systems.
\subsection{$q$-$k$-GM concurrence and $\alpha$-$k$-GM concurrence}
{\bf Definition 2}. For any $n$-partite pure state $|\psi\rangle$, the $q$-$k$-GM concurrence $(q>1,2\leq k\leq n)$ is defined as
\begin{equation}\label{9}
\begin{array}{rl}
\mathcal{G}_{q-k}(|\psi\rangle)=\frac{\big({\prod_{\alpha_i\in T_k}[2\sum_{t=1}^k(1-{\rm Tr}\rho_{A_{t{\alpha_i}}}^q)]^{1/2}}\big)^{1/|T_k|}}{\sqrt{k}},\\
\end{array}
\end{equation}
and the $\alpha$-$k$-GM concurrence $(0\leq\alpha<1,2\leq k\leq n)$ is given as
\begin{equation}\label{10}
\begin{array}{rl}
\mathcal{G}_{\alpha-k}(|\psi\rangle)=\frac{\big({\prod_{\alpha_i\in T_k}[2\sum_{t=1}^k({\rm Tr}\rho_{A_{t{\alpha_i}}}^\alpha-1)]^{1/2}}\big)^{1/|T_k|}}{\sqrt{k}}.\\
\end{array}
\end{equation}
Here $T_k=\{\alpha_i\}$ is the set constituted by all feasible $k$-partitions, $|T_k|$ represents the cardinality of the elements in the set $T_k$.

The above definition can be generalized to arbitrary general quantum states.

For any $n$-partite mixed state $\rho$, using convex-roof extension, we define $q$-$k$-GM concurrence as
\begin{equation}\label{11}
\begin{array}{rl}
\mathcal{G}_{q-k}(\rho)=\inf\limits_{\{p_i,|\psi_i\rangle\}}\sum_ip_i\mathcal{G}_{q-k}(|\psi_i\rangle),\\
\end{array}
\end{equation}
and $\alpha$-$k$-GM concurrence as
\begin{equation}\label{12}
\begin{array}{rl}
\mathcal{G}_{\alpha-k}(\rho)=\inf\limits_{\{p_i,|\psi_i\rangle\}}\sum_ip_i\mathcal{G}_{\alpha-k}(|\psi_i\rangle).\\
\end{array}
\end{equation}
Here the infimum runs over all possible ensemble decompositions of $\rho$.

Utilizing similar proof procedures to Theorem 1, we can show that $\mathcal{G}_{q-k}(\rho)$ and $\mathcal{G}_{\alpha-k}(\rho)$ obey necessary requirements including vanishing on any $\rho\in S_k$, being strictly positive for any $\rho\in S_1\backslash S_k$, entanglement monotone, strong monotone, convexity, thus we can draw the conclusion as follows.

{\bf Theorem 4}. Both $q$-$k$-GM concurrence and $\alpha$-$k$-GM concurrence are bona fide hierarchical entanglement measures.

It is obvious that the $q$-$k$-GM concurrence can be thought of as a generalization of $q$-concurrence \cite{2} or Tsallis-$q$ entanglement with a specific $q$ \cite{46} in terms of pure states, which is considered from different perspective than $q$-$k$-ME concurrence \cite{18}. Besides, $q$-$k$-GM concurrence can also be treated as the generalized form of genuine multipartite entanglement measure, GBC \cite{3} and geometric mean of $q$-concurrence $\mathcal{G}_{q}$ (G$q$C) \cite{41}.

Here we analyze two extreme cases. For any pure state $|\psi\rangle$, when $q\rightarrow\infty$, if $\mathcal{G}_{q-k}(|\psi\rangle)$ tends to $\sqrt{2}$, then $|\psi\rangle$ is $k$-nonseparable, if not, $|\psi\rangle$ is $k$-separable. In particular, if $\mathcal{G}_{q-2}(|\psi\rangle)$ tends to $\sqrt{2}$ as $q\rightarrow\infty$, then $|\psi\rangle$ is genuinely entangled state. However, if $\alpha=0$ and $|\psi\rangle$ is $k$-nonseparable, then
\begin{equation*}
\begin{array}{rl}
\mathcal{G}_{\alpha-k}(|\psi\rangle)=\frac{\big({\prod_{\alpha_i\in T_k}[2{\sum_{t=1}^k(r_{A_{t{\alpha_i}}}-1)}]^{1/2}}\big)^{1/|T_k|}}{k},
\end{array}
\end{equation*}
which means that $\mathcal{G}_{\alpha-k}(|\psi\rangle)$ relies on the rank $r_{A_{t{\alpha_i}}}$ of $\rho_{A_{t{\alpha_i}}}$. Therefore, this implies that $\mathcal{G}_{q-k}$ and $\mathcal{G}_{\alpha-k}$ characterize quantum states from different angles.

Furthermore, we will derive the lower bounds of these two classes of entanglement measures. First, we consider the relation between $q$-$k$-GM concurrence and $q$-$k$-ME concurrence.

{\bf Theorem 5}. The $\sqrt{2}$ times of $q$-$k$-ME concurrence is a lower bound of the $q$-$k$-GM concurrence.

{\bf Proof}. For any $n$-particle pure state $|\psi\rangle$, we see
\begin{equation*}
\begin{array}{rl}
\mathcal{G}_{q-k}(|\psi\rangle)&\geq\sqrt{2C_{q-k}(|\psi\rangle)}\\
&\geq\sqrt{2}{C_{q-k}(|\psi\rangle)},
\end{array}
\end{equation*}
where the first inequality can be gotten from the definition of $\mathcal{G}_{q-k}(|\psi\rangle)$ and $C_{q-k}(|\psi\rangle$, the second inequality is due to $[\sum_{t=1}^k1-{\rm Tr}\rho_{A_{t\alpha_i}}^q]/k\leq1$ of formula (\ref{21}).

For any mixed state $\rho$ with $\{p_i,|\psi_i\rangle\}$ being the optimal pure state decomposition of $\mathcal{G}_{q-k}(\rho)$, one derives
\begin{equation*}
\begin{array}{rl}
\mathcal{G}_{q-k}(\rho)&=\sum_ip_i\mathcal{G}_{q-k}(|\psi_i\rangle)\\
&\geq\sqrt{2}\sum_ip_iC_{q-k}(|\psi_i\rangle)\\
&\geq\sqrt{2}C_{q-k}(\rho),
\end{array}
\end{equation*}
where the last inequality is based on the definition of $C_{q-k}(\rho)$. $\hfill\blacksquare$

Considering the space of the PI quantum states instead of the entire space reduces the dimensionality of the considered space exponentially, so it is of utmost necessity to illustrate that $q$-$k$-GM concurrence and $\alpha$-$k$-GM concurrence meet the requirement given in Ref. \cite{5}.

{\bf Theorem 6}. The $q$-$k$-GM concurrence ($\alpha$-$k$-GM concurrence) is lower bounded by the the maximum of $q$-$k$-GM concurrence ($\alpha$-$k$-GM concurrence) of $\rho_{U}^{\rm PI}$, namely,
\begin{equation}\label{14}
\begin{array}{rl}
\mathcal{G}_{q-k}(\rho)\geq\max\limits_{U}\mathcal{G}_{q-k}(\rho_U^{\rm PI})
\end{array}
\end{equation}
and
\begin{equation}\label{15}
\begin{array}{rl}
\mathcal{G}_{\alpha-k}(\rho)\geq\max\limits_{U}\mathcal{G}_{\alpha-k}(\rho_U^{\rm PI}).
\end{array}
\end{equation}
Here $U$ is any locally unitary transformation.

The inequalities (\ref{14}) and (\ref{15}) can be directly gotten in a similar method to Theorem 3.

From the conclusions presented in this subsection, we can see that these two complete classes of generalized hierarchical entanglement measures still have the properties possessed by $k$-GM concurrence.
\subsection{The genuine entanglement measures}
It is well known that the genuinely entangled states are of great interest for several assignments, such as quantum phase transitions \cite{48}, teleportation and dense coding \cite{47}. As a result, it is of high necessity to give a reasonable quantification to determine whether a quantum state belongs to $S_1\backslash S_2$. Below we will present the special cases of $q$-$k$-GM concurrence and $\alpha$-$k$-GM concurrence corresponding to $k=2$, respectively.
\begin{figure}[htbp]
\centering
{\includegraphics[width=8.6cm,height=6cm]{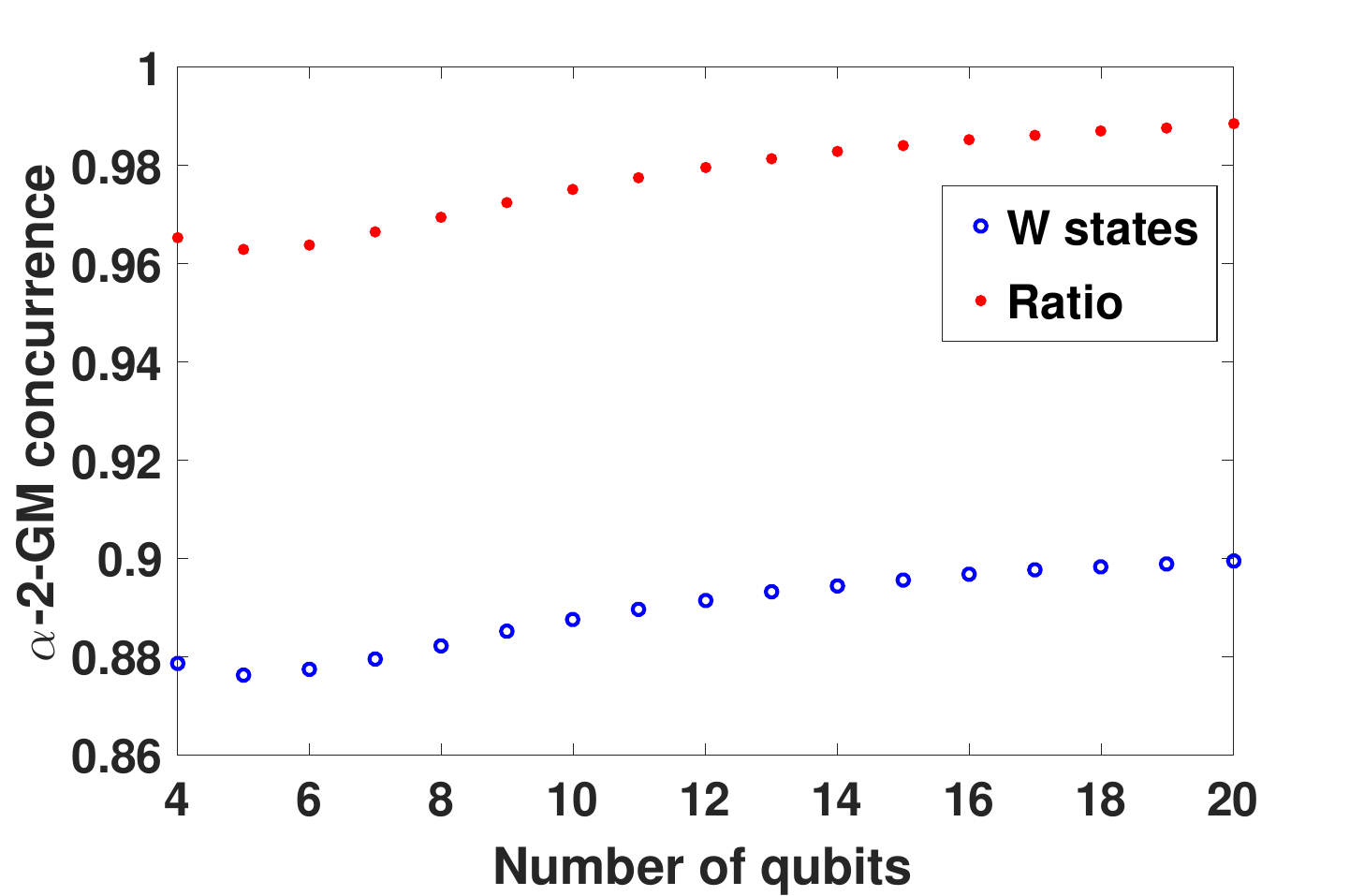}}
\caption{The blue hollow circles and the solid red circles are, respectively, the $\alpha$-2-GM concurrence of $n$-qubit $W$ state and the ratio of ${\mathcal{G}_{\alpha-2}(|W_n\rangle)}$ to ${{\mathcal{G}_{\alpha-2}(|{\rm GHZ}_n\rangle)}}$ when $\alpha=\frac{1}{2}$.\\}\label{fig 3}
\end{figure}

The $q$-$2$-GM concurrence $(q>1)$, for any $n$-partite pure state $|\psi\rangle$, can be written as
\begin{equation}\label{19}
\begin{array}{rl}
\mathcal{G}_{q-2}(|\psi\rangle)&=\big\{{\prod_{\alpha_i\in T_2}[2(1-{\rm Tr}\rho_{A_{t{\alpha_i}}}^q)]^{1/2}}\big\}^{1/|T_2|}.\\
\end{array}
\end{equation}
For any $n$-particle mixed state $\rho$, Eq. (\ref{11}) can be reduced as
\begin{equation}\label{20}
\begin{array}{rl}
\mathcal{G}_{q-2}(\rho)=\inf\limits_{\{p_i,|\psi_i\rangle\}}\sum_ip_i\mathcal{G}_{q-2}(|\psi_i\rangle).\\
\end{array}
\end{equation}

Another entanglement measure, $\alpha$-$2$-GM concurrence $(0\leq\alpha<1)$, is given by
\begin{equation}\label{16}
\begin{array}{rl}
\mathcal{G}_{\alpha-2}(|\psi\rangle)&=\big\{{\prod_{\alpha_i\in T_2}[2({\rm Tr}\rho_{A_{t{\alpha_i}}}^\alpha-1)]^{1/2}}\big\}^{1/|T_2|}\\
\end{array}
\end{equation}
for any $n$-partite pure state $|\psi\rangle$, and is expressed as
\begin{equation}\label{17}
\begin{array}{rl}
\mathcal{G}_{\alpha-2}(\rho)=\inf\limits_{\{p_i,|\psi_i\rangle\}}\sum_ip_i\mathcal{G}_{\alpha-2}(|\psi_i\rangle)\\
\end{array}
\end{equation}
for any mixed state $\rho$ of $n$-particle.

The $q$-$k$-GM concurrence ($\alpha$-$k$-GM concurrence) is the generalized form of $q$-$2$-GM concurrence ($\alpha$-$2$-GM concurrence), as such, $\mathcal{G}_{q-2}$ and $\mathcal{G}_{\alpha-2}$ respectively inherit the properties of $\mathcal{G}_{q-k}$ and $\mathcal{G}_{\alpha-k}$ automatically. Therefore, they can be regarded as two types of genuine entanglement measures which are able to separate the genuinely entangled states from the other quantum states.

It is acknowledged that there are two classes of genuine tripartite entangled states, the $W$ state and the GHZ state, which are not allowed to transform each other under stochastic LOCC \cite{43}. The capability of the $W$ state for quantum teleportation cannot exceed that of the GHZ state \cite{44}. As a result, the GHZ state is viewed as more entangled than the $W$ state.

The $n$-qubit GHZ state and $W$ state, respectively, can be denoted as
\begin{equation*}
\begin{array}{rl}
&|{\rm GHZ}_n\rangle=\frac{|0\rangle^{\otimes n}+|1\rangle^{\otimes n}}{\sqrt2},\\
&|W_n\rangle=\frac{|10\cdots0\rangle+|01\cdots0\rangle+\cdots|00\cdots1\rangle}{\sqrt n}.
\end{array}
\end{equation*}
The $q$-$2$-GM concurrence and G$q$C \cite{41} are two different kinds of entanglement measures for arbitrary quantum states, but in terms of pure states, one can obtain the relation $\mathcal{G}_{q-2}(|\psi\rangle)=\sqrt{2\mathcal{G}_{q}(|\psi\rangle)}$. By means of the results in Ref. \cite{41}, we can attain $q$-2-GM concurrence of $W$ state is strictly less than that of GHZ state in multiqubit systems.

Next, we will analyze whether $\alpha$-2-GM concurrence also obey the observation above.

The $\alpha$-concurrence of $|{\rm GHZ}_n\rangle$ and $|W_n\rangle$ are respectively
\begin{equation*}
\begin{array}{rl}
&C_{\alpha, p,n-p}(|{\rm GHZ}_n\rangle)=2^{1-\alpha}-1,\\
&C_{\alpha, p,n-p}(|W_n\rangle)=\frac{p^\alpha}{n^\alpha}+\frac{(n-p)^\alpha}{n^\alpha}-1,
\end{array}
\end{equation*}
where $C_{\alpha,p,n-p}$ stands for the $p$-to-other $\alpha$-concurrence.

\begin{widetext}
Then, one derives
\begin{equation*}
\begin{array}{rl}
&\mathcal{G}_{\alpha-2}(|{\rm GHZ}_n\rangle)=\sqrt{2(2^{1-\alpha}-1)},\\
&\mathcal{G}_{\alpha-2}(|W_n\rangle)=\begin{cases}\Big(\prod_{p=1}^{\frac{n-1}{2}}\big[\sqrt{2}\big(\frac{p^\alpha}{n^\alpha}+\frac{(n-p)^\alpha}{n^\alpha}-1\big)^{1/2}\big]^{C_n^p}\Big)^{1/|T_2|}, &n~{\rm is~odd},\\
\Big(\prod_{p=1}^{\frac{n-2}{2}}\big[\sqrt{2}\big(\frac{p^\alpha}{n^\alpha}+\frac{(n-p)^\alpha}{n^\alpha}-1\big)^{1/2}\big]^{C_n^p}\big(\sqrt{2(2^{1-\alpha}-1)}\big)^{{C_n^{\frac{n}{2}}}/{2}}\Big)^{1/|T_2|}, &n {\rm~is~even}.\end{cases}\\
\end{array}
\end{equation*}
When $\alpha=0$, $\mathcal{G}_{0-2}(|W_n\rangle)=\mathcal{G}_{0-2}(|{\rm GHZ}_n\rangle)=\sqrt{2}$, this means that $0$-$2$-GM concurrence does not fulfill the condition that GHZ state contains more entangled than $W$ state. When $\alpha$ belongs to the interval $(0,1)$, it is not difficult to prove that $C_{\alpha, p,n-p}(|{\rm GHZ}_n\rangle)\geq C_{\alpha, p,n-p}(|W_n\rangle)$ for any $p$ and $C_{\alpha, p,n-p}(|{\rm GHZ}_n\rangle)= C_{\alpha, p,n-p}(|W_n\rangle)$ holds when $p=\lfloor\frac{n}{2}\rfloor=\frac{n}{2},$ which implies $\mathcal{G}_{\alpha-2}(|{\rm GHZ}_n\rangle)>\mathcal{G}_{\alpha-2}(|W_n\rangle)$ for $n\geq3$.

The following we will elaborate that when $0<\alpha<1$, $\mathcal{G}_{\alpha-2}(|W_n\rangle)/{\mathcal{G}_{\alpha-2}(|{\rm GHZ}_n\rangle)}\rightarrow1$ as $n\rightarrow\infty$. The ratio of ${\mathcal{G}_{\alpha-2}(|W_n\rangle)}$ to ${{\mathcal{G}_{\alpha-2}(|{\rm GHZ}_n\rangle)}}$ is
\begin{equation*}
\begin{array}{rl}
\frac{\mathcal{G}_{\alpha-2}(|W_n\rangle)}{{\mathcal{G}_{\alpha-2}(|{\rm GHZ}_n\rangle)}}=\begin{cases}\frac{\Big(\prod_{p=1}^{\frac{n-1}{2}}\big[\big(\frac{p^\alpha}{n^\alpha}+\frac{(n-p)^\alpha}{n^\alpha}-1\big)^{1/2}\big]^{C_n^p}\Big)^{1/|T_2|}}{\sqrt{2^{1-\alpha}-1}}, &n~{\rm is~odd},\\
\frac{\Big(\prod_{p=1}^{\frac{n-2}{2}}\big[\big(\frac{p^\alpha}{n^\alpha}+\frac{(n-p)^\alpha}{n^\alpha}-1\big)^{1/2}\big]^{C_n^p}(\sqrt{2^{1-\alpha}-1})^{{C_n^{\frac{n}{2}}}/{2}}\Big)^{1/|T_2|}}{\sqrt{2^{1-\alpha}-1}}, &n {\rm~is~even}.\end{cases}\\
\end{array}
\end{equation*}
If $n$ is odd, then we set $n=2l+1$, one has
\begin{equation*}
\begin{array}{rl}
\lim\limits_{l\rightarrow\infty}\frac{\mathcal{G}_{\alpha-2}(|W_{2l+1}\rangle)}{{\mathcal{G}_{\alpha-2}(|{\rm GHZ}_{2l+1}\rangle)}}&\geq\lim\limits_{l\rightarrow\infty}\frac{\sum_{p=1}^lC_{2l+1}^p}{\sqrt{2^{1-\alpha}-1}\big(\sum_{p=1}^l C_{2l+1}^p\sqrt{\frac{(2l+1)^\alpha}{p^\alpha+(2l-p+1)^\alpha-(2l+1)^\alpha}}\big)}\\
&=\lim\limits_{l\rightarrow\infty}\frac{C_{2l+1}^l}{\sqrt{2^{1-\alpha}-1}C_{2l+1}^l\sqrt{\frac{(2l+1)^\alpha}{l^\alpha+(l+1)^\alpha-(2l+1)^\alpha}}}\\
&=1^-.
\end{array}
\end{equation*}
If $n$ is even, then we set $n=2l$, one gets
\begin{equation*}
\begin{array}{rl}
\lim\limits_{l\rightarrow\infty}\frac{\mathcal{G}_{\alpha-2}(|W_{2l}\rangle)}{{\mathcal{G}_{\alpha-2}(|{\rm GHZ}_{2l}\rangle)}}&\geq\lim\limits_{l\rightarrow\infty}\frac{\sum_{p=1}^{l-1}C_{2l}^p+\frac{1}{2}C_{2l}^l}{\sum_{p=1}^{l-1}C_{2l}^p\sqrt{\frac{(2l)^\alpha(2^{1-\alpha}-1)}{p^\alpha+(2l-p)^\alpha-(2l)^\alpha}}+\frac{1}{2}C_{2l}^{l}}\\
&=1^-.
\end{array}
\end{equation*}
\end{widetext}
Here the results are true following from the inequality $\frac{n}{\frac{1}{x_1}+\frac{1}{x_2}+\cdots+\frac{1}{x_n}}\leq\sqrt[n]{x_1x_2\cdots x_n}$ and Stolz-Ces\`{a}ro theorem \cite{42}.

Since $\mathcal{G}_{\alpha-2}(|{\rm GHZ}_n\rangle)>\mathcal{G}_{\alpha-2}(|W_n\rangle)$ for $n>2$, we obtain
\begin{equation*}
\begin{array}{rl}
\lim\limits_{n\rightarrow\infty}\frac{\mathcal{G}_{\alpha-2}(|W_{n}\rangle)}{{\mathcal{G}_{\alpha-2}(|{\rm GHZ}_{n}\rangle)}}=1^-.
\end{array}
\end{equation*}

Simultaneously, some concrete ratios of $\mathcal{G}_{\alpha-2}(|W_n\rangle)$ to ${\mathcal{G}_{\alpha-2}(|{\rm GHZ}_n\rangle)}$ are plotted in Fig. \ref{fig 3}, where $\alpha$ takes $\frac{1}{2}$ and $n$ ranges from 4 to 20. We can see intuitively that $\mathcal{G}_{\alpha-2}(|W_n\rangle)/{\mathcal{G}_{\alpha-2}(|{\rm GHZ}_n\rangle)}$ tends to 1 as $n$ increases, but it is always less than 1.

\section{Conclusion}\label{VII}
In this work, we propose and discuss three full categories of hierarchical entanglement measures, $k$-GM concurrence, $q$-$k$-GM concurrence $(q>1)$, and $\alpha$-$k$-GM concurrence $(0\leq\alpha<1)$, which can be used as indicators to accurately identify $k$-separability $(2\leq k\leq n)$ of $n$-partite quantum states. In addition, they can not only finely classify entangled states into ($n-1$) different graduation in accordance with degrees of inseparability, but also discriminate some quantum states in the same hierarchy that cannot be identified by $k$-ME concurrence. In particular, $k$-GM concurrence covers the GBC \cite{3} and is a special case of $q$-$k$-GM concurrence corresponding to $q=2$. The $k$-GM concurrence meets the axiomatic requirements of an entanglement measure and exhibits differences in measuring continuity and entanglement orders when compared to $k$-ME concurrence. Moreover, we establish a relation between $k$-GM concurrence and $k$-ME concurrence, and verify that $k$-GM concurrence conforms to the requirement that $k$-GM concurrence of the PI part of quantum state $\rho$ is a lower bound on $k$-GM concurrence of original state. Furthermore, we illustrate that the parametrized generalized hierarchical measures, $q$-$k$-GM concurrence and $\alpha$-$k$-GM concurrence, also have the properties owned by $k$-GM concurrence, in order to describe the entanglement characteristics of states from multiple dimensions. Besides, $\alpha$-2-GM concurrence $\mathcal{G}_{\alpha-2}$ $(0<\alpha<1)$ is shown to satisfy that GHZ state contains more entanglement than $W$ state in multiqubit systems, and we obtain that ${\mathcal{G}_{\alpha-2}(|W_n\rangle)}/{{\mathcal{G}_{\alpha-2}(|{\rm GHZ}_n\rangle)}}$ tends to 1 as the number of particles approaches infinity but is always less than 1. Our study may have important implications for understanding the hierarchies of multipartite entanglement.

\section*{ACKNOWLEDGMENTS}
This work was supported by the National Natural Science Foundation of China under Grants No. 12071110 and No. 62271189, and the Hebei Central Guidance on Local Science and Technology Development Foundation of China under Grant No. 236Z7604G.


\end{document}